# Magnetization switching driven by spin-transfer-torque in high-TMR magnetic tunnel junctions.

D. Aurélio[a*], L. Torres[a] and G. Finocchio[b]

[a] Departamento de Física Aplicada. Universidad de Salamanca. Plaza de la Merced s/n 37008 Salamanca, Spain.

[b] Dipartimento di Fisica della Materia e Ingegneria Elettronica. University of Messina. Salaita Sperone 31, 98166 Messina, Italy.

**Abstract**

This paper describes a numerical experiment of magnetization switching driven by spin-polarized current in high-TMR magnetic tunnel junctions (TMR>100%). Differently from other works, the current density distribution throughout the cross-sectional area of the free-layer is here computed dinamically, by modeling the ferromagnet/insulator/ferromagnet trilayer as a series of parallel resistances. The validity of the main postulated hypothesis, which states that the current density vector is perpendicular to the sample plane, has been verified by numerically solving the Poisson equation. Our results show that the nonuniform current density distribution is a source of asymmetry for the switching. Furthermore, we found out that the switching processes are characterized by well defined localized pre-switching oscillation modes.

---

[*] Corresponding author
E-mail address: davidaurelio@usal.es

# 1. Introduction

The theoretical prediction [1,2] and experimental validation [3-10] that a spin-polarized current flowing trough a nanomagnet can alter its magnetic state, opened a whole range of perspectives for both theoretical and applied physics. In fact, a deep understanding of spin transfer torque theory should help in optimizing the design of spintronic devices like, nano-oscillators, [10] radio-frequency detectors, [11] and magnetoresistive random access memories (MRAM) [12]. In particular, concerning storage applications, magnetic tunnel junctions (MTJs) are the most promising candidates as a basis for an MRAM cell, this thanks to the techniques that have been developed to make tunnel barriers (i.e. $Al_2O_3$, [8] MgO, [13] and organic insulator [14]) sufficiently thin in order to support the necessary current densities, that produce magnetic switching with spin transfer torques, maintaining very large room temperature tunnel magnetoresistance (TMR). In this last aspect, state of art MTJs for MgO tunnel barriers have been reported to surpass 500% (TMR) at room temperature, [15] with critical current densities of the order of $10^6$ A/cm$^2$.

Since the switching processes can occur via nucleation processes, in particular for collinear configurations as measured by x-ray microscopy, [16], it is thus necessary to make use of micromagnetic computations in order to reproduce or predict the experimental behaviour [17]. Generally, in these computations the current density distribution is considered as being uniformly distributed throughout the entire cross-sectional area of the device, however this approximation is not realistic especially when considering high-TMR MTJs. For example, considering the simple scenario where a 180° spatial magnetization domain configuration is nucleated, in the region where the magnetization is parallel (P) to the reference magnetization of the pinned layer (PL) the resistance (current density) is smaller (larger) compared to that of the anti-parallel (AP) region.

In this work, we report micromagnetic simulations of the magnetization switching in high-TMR MTJs taking into account nonuniform current density (NUCD) distribution implemented by means of a parallel resistance channel model, where the main hypothesis postulated is that the current density vector is perpendicular to the sample plane.

The voltage dependence of the TMR has not been considered into our simulations, thus overestimating the nonuniform current distribution effect, nevertheless by comparing our results with numerical computations using the uniform current density (UCD) distribution, gives the possibility to fix a working limit to the switching behavior. In other words, the real spatial distribution of the current density will be between the uniform and the nonuniform distributions obtained without taking into account the voltage dependence of the TMR.

We have studied nanopillars with an elliptical cross-sectional area of CoFe (8nm) / MgO (0.8nm) / Py (4nm), with major plus minor axes of 90 nm and 35 nm respectively, the CoFe is exchange biased and acts as the PL, while the Permalloy (Py), represents the free layer (FL).

The paper is organized as follows: section 2 describes the numerical details of the micromagnetic computations; sections 3 and 4 show the results and their discussion, plus the conclusions of our numerical experiments.

## 2. Theory

The magnetoresistance $R$ of the MTJ depends on the relative orientation of the normalized magnetization of both the FL ($\vec{m}$) and the PL ($\vec{p}$) and can be approximated within the macrospin scope by

$$R = R_P + \frac{\Delta R}{2}\left[1 - \left(\vec{m}(x,y,z)\cdot\vec{p}(x,y,z)\right)\right]$$ where $\Delta R = R_{AP} - R_P$, with $R_{AP}$ and $R_P$ as the resistances of the AP and P state respectively.

By assuming that the current density vector is perpendicular to the sample plane $\vec{J} = J_z(x,y)\vec{u}_z$, and introducing a 2D numerical discretization in order to describe the magnetic system within the micromagnetic framework (we consider the $\vec{p}$ pinned along the $+x$), the $J_z(x,y)$ is computed for each computational cell thus becoming a state-dependent function $J_z(x,y,\vec{m})$, in other words it is a nonlocal term since it depends on the spatial configuration of the magnetization in the whole FL. The FL is discretized in cells of dimensions 2.5×2.5×4 nm³ and area $\Delta S = 6.25 \text{ nm}^2$. Let $N$ be the total number of cells, the current is then given by

$$I_0 = \Delta S \sum_{i,j} J_z(i,j,\vec{m}) = \frac{S}{N}\sum_{i,j} J_z(i,j,\vec{m})$$, where $S$ is the cross-sectional area of the FL. We use two indexes $(i,j)$ in order to distinguish between each orientation of the two dimensional discretization of the FL. Considering $N$ parallel channels (see Fig. 1) of resistance $r(i,j) = \frac{SR_P}{\Delta S} + \frac{S\Delta R}{2\Delta S}\left[1 - \left(\vec{m}(i,j)\cdot\vec{p}\right)\right]$ and considering that

$J_0 = \dfrac{I_0}{S}$, we can compute the current density distribution for each channel using simple circuit theory considerations:

$$J_z(i,j,\vec{m}) = \dfrac{J_0}{r(i,j)\sum_{i,j}\dfrac{1}{r(i,j)}} \qquad (1)$$

which represents the final expression of the time dependent NUCD distribution implemented in our micromagnetic framework. The Landau-Lifshitz-Gilbert-Slonczewski LLGS [18] equation thus becomes:

$$(1+\alpha^2)\dfrac{d\vec{m}}{d\tau} = -\vec{m}\times\vec{h}_{eff} - \alpha\,\vec{m}\times(\vec{m}\times\vec{h}_{eff}) - \dfrac{g\,\mu_B J_z(\vec{m})}{L_z\gamma_0 M_S^2 e}g_T(\vec{m},\vec{p})\left(\vec{m}\times(\vec{m}\times\vec{p}) - \alpha\,\vec{m}\times\vec{p}\right) \qquad (2)$$

where $g$ is the gyromagnetic splitting factor, $\gamma_0$ is the gyromagnetic ratio, $\mu_B$ is the Bohr magneton, $\alpha$ is the damping parameter, $L_z$ is the thickness of the FL, $e$ is the electron charge, $d\tau = \gamma_0 M_S dt$ represents the dimensionless time step, and $M_S$ is the saturation magnetization of the FL. The effective field $\vec{h}_{eff}$ takes into account the standard micromagnetic contributions (external, anisotropy, self-demagnetizing and exchange fields), the magnetostatic coupling between PL and FL, and the Oersted field; the latter, differently from previous studies [19,20], depends on the magnetization since $\vec{h}_{Oe} = \vec{h}_{Oe}(J_z(m))$.

For the scalar function $g_T$ we used the expression proposed by Slonczewski in 2005, rewritten in terms of current, $g_T(\theta) = 0.5\eta_T\left[1+\eta_T^2\cos(\theta)\right]^{-1}$ [21], where $\cos(\theta) = \vec{m}\cdot\vec{p}$ and $\eta_T$ is the polarizing factor [22,23]. The numerical parameters used in the simulations were: an external field of $H$=50 mT (to compensate the magnetostatic coupling with the PL), $M_S$=6.44×10$^5$A/m, $M_{SP}$=1.15×10$^6$A/m (saturation magnetization of the pinned layer), $\alpha$=0.01, $\eta_T$=0.7, [24] an exchange constant of $A$=1.3×10$^{-11}$ J/m, $R_P$=100 Ω, and $R_{AP}$=200 Ω, which are typical experimental parameters [25,26].

In order check the validity of our main hypothesis (current density vector perpendicular to the sample plane) a commercial finite element software (MagNet) was used to calculate the spatial current density distribution. This software uses a 3D drawing worksheet that allows us to draw the 3D MTJ geometry, from which the software automatically generates the finite element mesh. The different conductivities of each layer are introduced in order to perform the computation, (see Fig. 1 and Fig. 2; $\sigma_{Cu}$=5.77×10$^7$ S/m; $\sigma_{AF}$=$\sigma_{PL}$=$\sigma_{MgO\ barrier}$=1.111×10$^3$ S/m; $\sigma_{CoFe+8\%}$=1.029×10$^3$ S/m; $\sigma_{CoFe-16\%}$=1.322×10$^3$ S/m). To test our previously mention hypothesis, we introduce a variation in the conductivity of the FL, decreasing it in the central area by 16% and increasing it in the outer regions by 8% of the nominal value, which would correspond to a 360º domain configuration. The spatial distribution of the conductivity can be visualized in Fig. 2(a) (the lighter central area represents where the conductivity was decreased by 16%, and the darker lateral regions increased by 8%). After performing the computation the spatial distribution of the current density was evaluated.

Analyzing the results of the MagNet's simulation (Fig. 2) we see that the current density vector $\vec{J}$ mainly lays in the normal direction (z), representing in the worst case ~95% of the total $\vec{J}$. Looking at the variation of the normal component of $\vec{J}$ ($J_n$ in the Fig. 2(a)), one can see that in the FL there are two distinct zones, each corresponding to a different value of the conductivity, where the most significant variations of $J_n$ are observed at the border of those two regions. Crosschecking this result by plotting the tangential component of $\vec{J}$ ($J_t$ in the Fig. 2(b)) it is possible to see a gradient of the current density towards the border of the conductivity variation (equivalent to a domain wall), nevertheless $J_t$ is always less than 2% of $J_n$, thus confirming our approximation that $\vec{J}$ is perpendicular to the sample plane as can be seen in Fig. 2(c).

## 3. Results and discussion

Having our group previously carried out micromagnetic studies of some aspects of the magnetization switching [17] and persistent dynamics [26,27] in MTJs using an UCD distribution micromagnetic model, we will here point out the effects of the NUCD distribution model in numerical simulations of the switching process considering $R_{AP}$=200Ω and $R_P$=100Ω (simulations performed considering $R_{AP}$=250Ω gave qualitatively the same results), and compare it with the previous model by two different approaches. The first will be focused on current density pulsed excitations through the nanopillar, and the second in studying the effect of continuing injecting an increasing current density through the device for both transitions (P→AP and AP→P).

3.1 Magnetization switching with a current pulse: Modal analysis and phase diagrams

This section presents a detailed study of the magnetization switching by means of current pulses for both models. This kind of study is important from either the technological (writing mechanism in MRAM) or basic physical point of view, since it provides information about the stability of intermediate states and the way in which energy is pumped into the system.

An example of a magnetization switch is shown in Fig. 3(a) where the x-component of the normalized average magnetization is displayed, for a pulse of amplitude $J_0$=4.5×10$^6$ A/cm$^2$ applied during 14 ns with rising and descending times of 0.1 ns. As a first analysis between models we observe that the NUCD promotes the AP→P transition and that for both models the switching process begins by means of preliminary oscillations. In order to study how these pre-switching oscillations might influence the magnetization spatial distribution a detailed inspection within the frequency domain was performed by applying a Micromagnetic Spectral Mapping Technique (MSMT) [28,29]. This technique, applied from the initial instant up to 12.8ns, will allow us to compute the excited pre-switching oscillation modes. Looking at the frequency domain spectra shown in Fig. 3(b) one sees that the main mode ($f$=6.5 GHz) is basically the same for both the UCD and NUCD models, however there are differences in the lower frequency modes, existing more modes and with larger power in the NUCD case. To have a better analysis of the modes we have also computed 2D density plots (spatial distribution) of the power intensity of the excited modes [28] (insets of Fig. 3 and 4b). This will give us information of which parts of the sample oscillate for a given mode. Figure 3(b) shows two types of pre-switching oscillation modes in the AP→P transition for both models. The main mode is localized in the lateral areas of the sample, which we define as an "edge" mode, while the low power modes are in the central area of the sample, and thus defined as "central" modes. In terms of the spatial distribution of the oscillation modes (insets of Fig. 3(b)), we can see that the switching process is very similar between both models, however the NUCD model presents one more central mode than the UCD one. The presence of this extra central mode, and the fact that these secondary modes are in general more intense for the NUCD model, may be the underling cause for the faster transition in the NUCD case. In this case the increase of power in the central modes could be explained as follows: when the magnetization in the central region begins to oscillate the resistance starts to decrease leading to a current density increase in that same area, therefore the spin torque is augmented and the oscillations promoted. From another point of view, the oscillations begin in the edges (see main mode at 6.5 GHz in Fig. 3(b), and inset (v) of Fig. 6), but before the switching takes place, the symmetry has to be broken, this implies that the central region has to be destabilized from its static configuration along the easy axis which for the NUCD case happens faster (for this transition) because the current is focused in the edges of the FL at the initial moments promoting the oscillations.

The P→AP switching has also been analyzed with the MSMT (see Fig. 4 ($J_0$=1.05×10$^7$ A/cm$^2$)), whose results show a main pre-switching mode (6.5 GHz) equal to the AP→P transition and different edge modes around 5.5 GHz and 5.8GHz (NUCD). Comparing both models for this transition, the switching process is also practically the same with similar pre-switching oscillation modes, but with the difference that now is the UCD distribution that presents an extra oscillation mode and switches a bit faster. This effect is due to the distribution of the current density in the NUCD model that for this transition promotes less oscillation modes. As in the previous transition the oscillations start in the edges, but here it makes the resistance higher in those regions, thus during the initial moments the current is focused in the center of the FL where the oscillations are more damped, and as a consequence the switching is delayed.

In general it seems that the switching comes faster when more low frequency oscillation modes are excited (since the main high frequency one remains the same for both models), where the appearance of more or less modes for the NUCD model can be explained by the current density distribution throughout the FL.

To further extend the analysis of current density pulsed assisted switching, a systematic study was performed in order to compile comparative phase diagrams (see Fig. 5) as function of the height ('y'-axis) and duration ('x'-axis) of the current density pulse. The current was increased linearly from zero, up to its maximum value in 100 ps (and a resolution of 3×10$^5$ A/cm$^2$ for the current density, was considered in all phase diagrams).

As a general trend from Fig. 5, we observe that for the NUCD model the boundary between switched and nonswitched states is smoother, without much of the "teeth" behavior seen when using the UCD model. Nevertheless, a fully smooth frontier is not obtained for the NUCD model. For example in the P→AP transition with the external field perfectly aligned with the easy axis and for current density pulses of 1.14×10$^7$ A/cm$^2$, a "teeth" region is present. A detailed analysis of the switching in that region shows that the intermediate states of the switching are extremely inhomogeneous, including cases of vortex and anti-vortex configurations (not showed) [16]. It is due to these configurations, influenced by intrinsically nonuniform effective fields like the Oersted field, that although the current density is increased for a certain pulse, the switching event is not achieved, even though it had been achieved for smaller current densities (black "teeth" in Fig. 5).

By comparing both models (Fig. 5) we can say that the NUCD distribution is a more stable model given that it is not as propitious as the UCD model to form "teeth" structures. On these diagrams it is also seen that the NUCD favors the AP→P and hinders the P→AP as it was seen in the particular case of figures 3 and 4. In particular for P→AP transition the current density needed to achieve the switching, when applying the NUCD model, is significantly larger than when using the UCD model, this due to the damped oscillations as a result of focusing the current density in the central area of the FL, as it was more thoroughly described above for Fig. 4. This makes that in terms of critical density current the NUCD model is more asymmetric that the UCD one, and since all parameters are the same for both models, the NUCD distribution could be given as a possible contribution to the intrinsic asymmetry of the switching seen in real devices.

Since former work in this kind of nanostructures, both experimental [30,31] and theoretical [17], demonstrated that a small misalignment of the external field regarding the easy axis gives rise to more uniform magnetization dynamics and lower critical currents, phase diagrams where the external field was tilted 3º out of the easy axis were also computed. As expected, one can see (Fig. 5) how the critical current densities decrease and the frontiers become smoother as the result of a more uniform behavior of the magnetization dynamics, this due to the hard axis component of the external field whose torque helps in pulling the averaged magnetization off the easy axis equilibrium position.

3.2 Ramped current hysteresis loops

As previously mentioned we also performed hysteresis loop studies by continuingly increasing the current density injected through the nanopilar until the switching critical value, in order to see how would both models behave in this regime.

Figure 6 shows ramped current density magnetoresistance hysteresis loops computed using the UCD and NUCD models. Looking at the hysteresis loops one can see that NUCD model favors the switching going from the AP→P state, and on the contrary it hinders the reverse transition P→AP, when compared to the UCD distribution, which is in agreement with the discussion seen in the previous section but here compared in terms of critical current density.

Analyzing both switching processes (for a sweep rate of $10^{13}$ A/(cm$^2$ s)), again we see that they are preceded by initial oscillations of the FL magnetization, which we will now describe. In insets (ii) for the P→AP transition and (v) for the AP→P case, one can see that the non-uniform oscillations start and are focused at the lateral regions. As the amplitude of these oscillations increase, they break the magnetization symmetry and lead to several complex states (insets (iii) and (vi)) before the switching is achieved.

Focusing now on the introduction of the NUCD model, we analyze how it leads to opposite effects depending on the transition analyzed. In the AP→P transition, the oscillations at the lateral regions produce a decrease in resistance at those same regions, thus the application of the NUCD model results in a higher current injection in those lateral areas (inset (iv) of Fig. 6), promoting larger oscillations and triggering the switching. On the other hand, for the P→AP reversal the oscillations at the outer regions imply an increase of the resistance there and thus less current is injected in that region (inset (i) Fig. 6), hence the oscillations are damped and the switching delayed in terms of critical density current. Of course this is just a simplified description of the process and indeed the magnetization dynamics are rather nonlinear. We have also plotted the critical current densities as a function of the sweep rate (Fig. 7). Here we see that these curves are nonmonotonic, and as reported in other numerical studies [19,20] its trend is related to a trade-off between the Oersted field and the spin-transfer torque. Although the observed behaviour is not linear, the general trends of applying the NUCD model follow the aforementioned discussion seen in Fig.6.

Figure 7 also shows the frequency of the main (larger power) pre-switching oscillation mode as a function of the sweep rate, for each transition and model. Again depending on the transition analyzed, different results are observed. In the AP→P transition (Fig. 7(a)) the critical current density and frequency of the main mode in function of the sweep rate, are more or less independent of the spatial current density distribution. On the other hand the critical current density and frequency of the main mode in the P→AP transition are clearly affected by the use of the NUCD. In order to better understand the physical mechanisms that originate this behavior, we performed a modal analysis using the MSMT (see Fig. 8) similar to the ones showed in figures 3 and 4. Nevertheless, we stress that a direct comparison with the pulsed current case is not possible. The current is applied linearly (with a slope equal to SR) implying that when the switching is achieved the spin torque has been acting and pumping energy for a much longer time. For example, in the AP→P case, for the slowest sweep rate (Fig. 8(a), SR = $5\times10^{12}$ A/cm$^2$.s) when the switching is achieved the current has been pumping energy for 610 ns. For the same transition in the pulsed case analyzed in Fig. 3, the current applied up to the switching point was of 13 ns. Subsequently we will comment in higher detail the modes excited in both transitions.

In the AP→P transition (Fig. 7(a)), as previously stated, the critical current density and frequency of the main mode in function of the sweep rate are practically independent of the spatial distribution of current density. This occurs because the excited modes are the same for both UCD and NUCD distributions, giving rise to practically identical nucleation processes for the switching (see Figs. 8(a) and 8(b)).

For the slowest sweep rate (Fig. 8(a)) the main mode is a center lower frequency one, in accord with the slower pumping of energy. On the other hand for the fastest SR analyzed (Fig. 8(b) SR=$10^{14}$ A/cm$^2$.s) the edge mode is the leading one, corresponding to the faster way of pumping energy. In other words, the difference in the excitation velocity (slope of the current density ramp, i.e. SR) leads to a different formation of modes in the sample. Between the minimum and maximum SR an intermediate behavior is found, with a minimum in the critical current density where both edge and center modes have similar importance. The jump in frequency after this minimum in critical current density $J_{min}$ (SR=$1.67\times10^{13}$ A/(cm$^2$ s)) reveals a transition from the edge predominant mode dynamics to the center predominant mode dynamics. We also observe that this mechanism is related to a trade off between the Oersted field and the spin-torque, given that by removing the Oersted field effect from the simulations the minimum is not present. This happens because when the Oersted field is removed, the predominant mode is always a center one around 4 GHz (similar to the one of Fig 8(a)). This is expected since the nonuniform Oersted field (more intense near the boundary of the sample) clearly promotes the edge modes.

Finally it is worth to comment that for the studied AP→P transitions, even though the NUCD model slightly diminishes both the switching time (seen in the pulsed cases of Figs 3 and 5) and critical current density (seen in the ramped current loops of Figs. 6 to 8), the effect is, in both cases, very weak. Subsequently, we can conclude that the NUCD model weakly affects the AP→P transition.

Analyzing now the P→AP reversal (Fig. 7(b)), one sees that the frequency of the main mode is clearly different between the UCD and NUCD models but, in a good approximation, independent of the sweep rate for both models. This can be understood by looking at the modal analysis of Figs 8(c) and 8(d), where one can see that for the UCD model the predominant mode is a center one for all the SR, while for the NUCD the predominant mode in the dynamics is an edge one. Again, as expected due to the way in which the energy is pumped into the system, the center modes are more intense for slower SR, while the edge modes gain intensity for faster SR, nevertheless the predominance of one or the other does not change in the simulated SR range. The predominance in this case depends mainly on the use of the NUCD model. We also observed that in this case the dynamics is not affected at all by the suppression of the Oersted field pointing to a more spin torque dependent excitation of the modes.

Comparing both models in terms of critical current densities for these transitions (P→AP), we see that a higher current density is needed to switch the system in the NUCD model, which is in agreement with the previous section discussion and the naive one based on Fig. 6.

Summing up the results of both models and transitions, we see that the P→AP transition is more affected by the use of the NUCD distribution. For the NUCD model, bringing together the naive discussions of Fig. 6 and all the modal analysis performed, it could be stated that after the initial oscillations in the edges (due mainly to the magnetostatic coupling with the fixed layer) the concentration of the current in the central area does not significantly promote the center modes (P→AP case). The main effect of this concentration of the current density in the center seems to be a frequency reduction of the edge mode in respect to the AP→P transition (Compare Figs. 3 and 4 or 8(a) and 8(b) with 8(c) and 8(d)).

Since the effect of the NUCD model is clearly different for AP→P and P→AP transitions, in real devices the effect of the NUCD distribution could therefore be taken as an additional source for the intrinsic asymmetry in the switching process.

## 4. Conclusion

In summary, we have performed micromagnetic studies of the magnetization switching driven by spin-polarized current in a high-TMR MTJ, studying the effects of two current density distribution models one uniform and the other nonuniform using, an increasingly current density up to the transition point and current density pulsed excitations.

From our results we see that the NUCD distribution is a source of asymmetry between both transitions in terms of critical current density, verified in both the pulsed and ramped excitations. This asymmetry is highlighted in the numerical experiments using an increasingly current density ramped excitation, which showed that the AP→P switching is only marginally affected by the current density distribution while its effect on the P→AP transition is much more significant, in both critical current density and generated oscillation modes.

In the analysis within the frequency domain, for current density pulsed excitations, we see that the pre-switching oscillations are characterized by "edge" and "central" modes, and accordingly to the model used more or less modes are excited, coming the transition faster when more oscillation modes are generated. These types of modes have been recently experimentally detected [32]. For ramped current hysteresis loops the modal analysis also shows that the effect on critical current can be explained based on the predominance of these center or edge modes within the dynamics.

Making an extensive study using both models in the pulsed excitation regime, we get that in general the NUCD distribution seems to be a more uniform model, having smoother transitions in a current density versus pulse duration diagram (Fig.5), showing as well the aforementioned higher asymmetry between transitions, and that a small misalignment of the external field in respect to the easy axis reduces the critical current density.

The approach used in this paper of the NUCD distribution can also be generalized to micromagnetically describe the presence of pinholes in the tunnel barrier, in particular when the conductive channel presents a resistance comparable to the ideal MTJ [33].

**Acknowledgments**

The authors thank Victor Raposo for assistance in the MagNet software and helpful discussions. This work was partially supported by Spanish projects MAT2008-04706/NAN and SA025A08.

———————————————————

FIG. 1 (Color online) Sketch of the MTJ we simulated and its approximation with the parallel channel resistance model..

FIG. 2 (Color online) Results of the current densities computed using MagNet (color gradient represents ($J$ in A/m$^2$)): (a) spatial distribution of the normal component of $J$ in the FL. (b) spatial distribution of the tangential component of $J$ in the FL. (c) Horizontal cross-section of the FL In which the arrows represent the current density vector (d) zoom of the circle in Fig. 2. (c).

FIG. 3 (Color online) AP→P transition. (a) Normalized 'x' component of the magnetization versus time for uniform (gray line) and NUCD model (color line). The height of the current density pulse applied was of 4.5×10$^6$ A/cm$^2$, with a pulse duration of 14ns and rising and descending times of 100ps. (b) Frequency spectrum for the UCD (gray thin line) and NUCD model (color thick line) of the pre-switching oscillations from 0 to 12,8 ns. Insets: Power density 2D plots produced by each computational cell at the FL with the corresponding frequency mode indicated for each model, (darker means larger power).

FIG. 4 (Color online) P→AP transition (a) Normalized 'x' component of the magnetization versus time for uniform (gray line) and NUCD model (color line). The height of the current density pulse applied was of -1.05×10$^7$ A/cm$^2$, with a pulse duration of 14ns and rising and descending times of 100ps. (b) Frequency spectrum for the UCD (gray thin line) and NUCD model (color thick line) of the pre-switching oscillations from 0 to 12,0 ns. Insets: Power density 2D plots produced by each computational cell at the FL with the corresponding frequency mode indicated for each model, (darker means larger power).

FIG. 5 (Color online) Phase diagrams of the current density pulsed excitation switching for both the AP→P (top) and P→AP (bottom) transitions. From left to right, NUCD and UCD model with a *50* mT external field applied along the easy axis, with the last NUCD diagram where the external field applied as a 3º misalignment in respect to the easy axis. Color area means that the system has switched, and black the opposite.

FIG. 6 (Color online) Magnetization vs critical current density hysteresis loops. In color the results using the NUCD and in black using UCD distribution for a sweep rate 10$^{13}$ A/(cm s). Insets (i) and (iv) represent the spatial distribution of the current density (see the coordinate in the figure). Insets (ii) and (v) show the magnetization snapshot at the beginning of the magnetization switching (iii) and (vi) show the magnetization snapshot before the switching is achieved.

FIG. 7 (Color online) Both AP→P, (a), and P→AP, (b), reversals comparing the critical current density between the NUCD model (color line) and the uniform one (black line), in function of the sweep rate. On the right scale of each graph, it is represented the main pre-switching oscillation frequency mode for both models and transitions at each sweep rate tested, (doted line).

FIG. 8 (Color online) Frequency spectrum for both models and transitions in respect to the lowest and highest sweep rates (SR) of Fig. 7. (a) and (b) Pre-switching oscillation modes of both AP→P transitions, show that the modes generated are equal between models and that for lower sweep rates (lower than the one of minimum critical current density of Fig. 7 (a)) the main mode turns in to a central one (see insets). (c) and (d) Pre-switching oscillation modes of the P→AP transitions, show for the UCD model more oscillation modes are generated being the main one central, while for the NUCD the main mode is an edge one, (cause of the difference in frequency seen in Fig.7(b)).

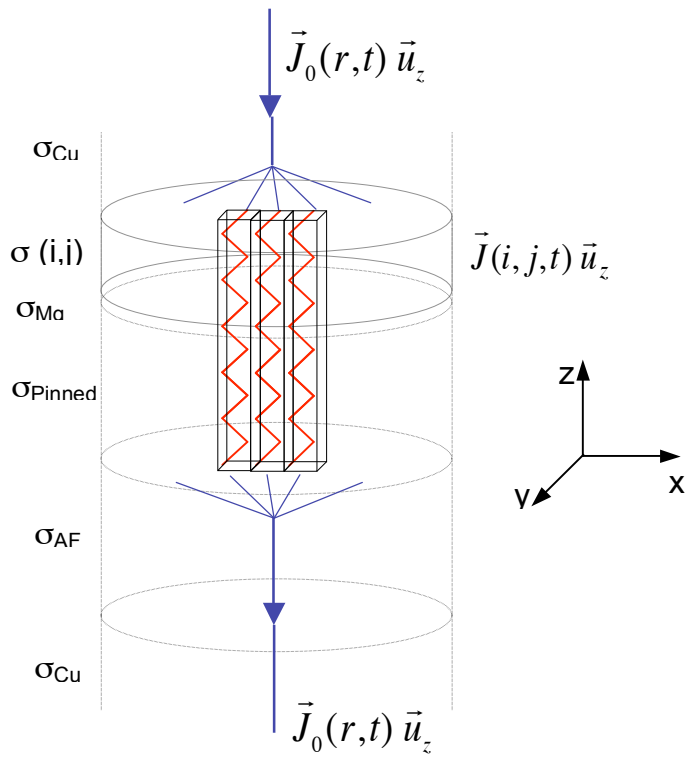

Fig. 1

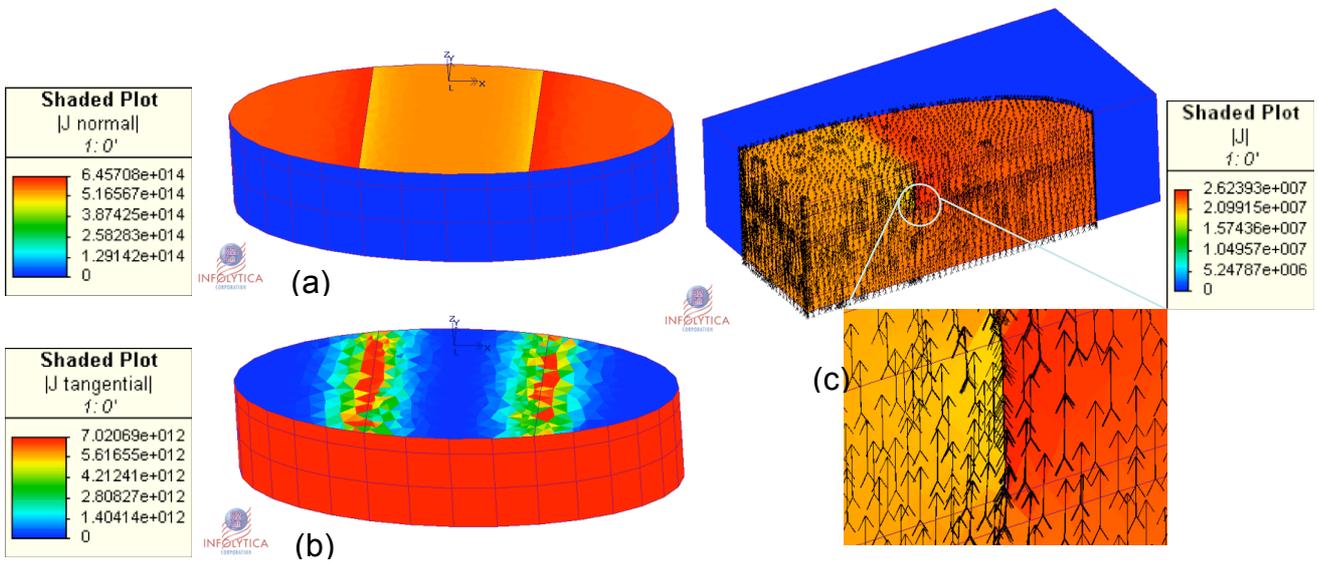

Fig. 2

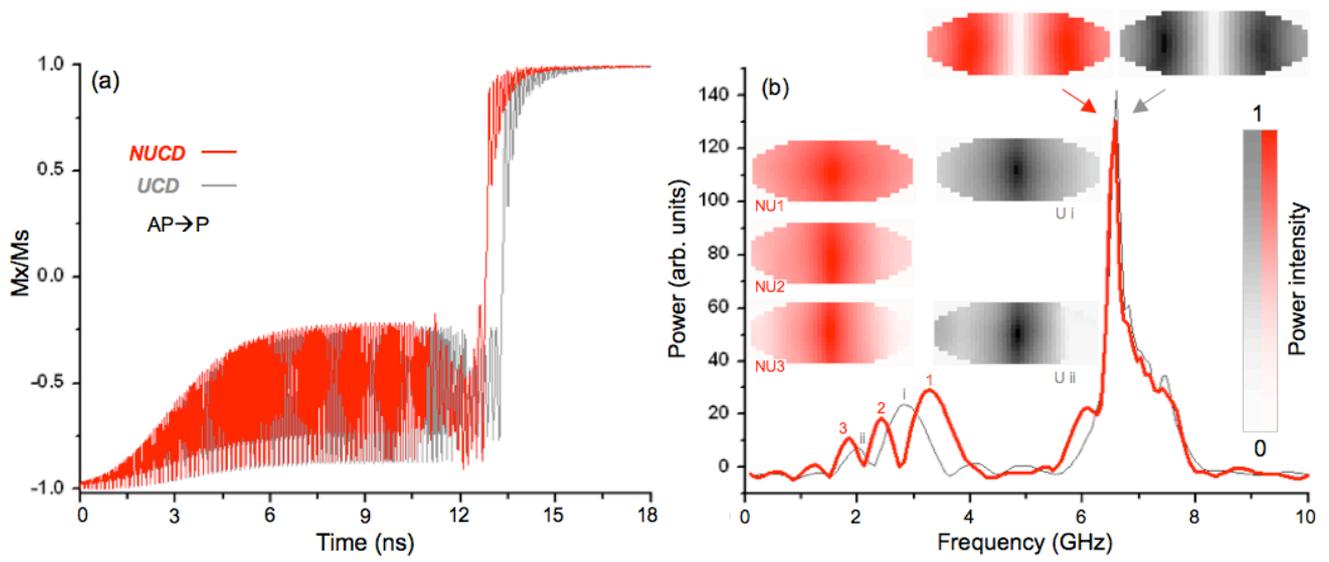

Fig. 3

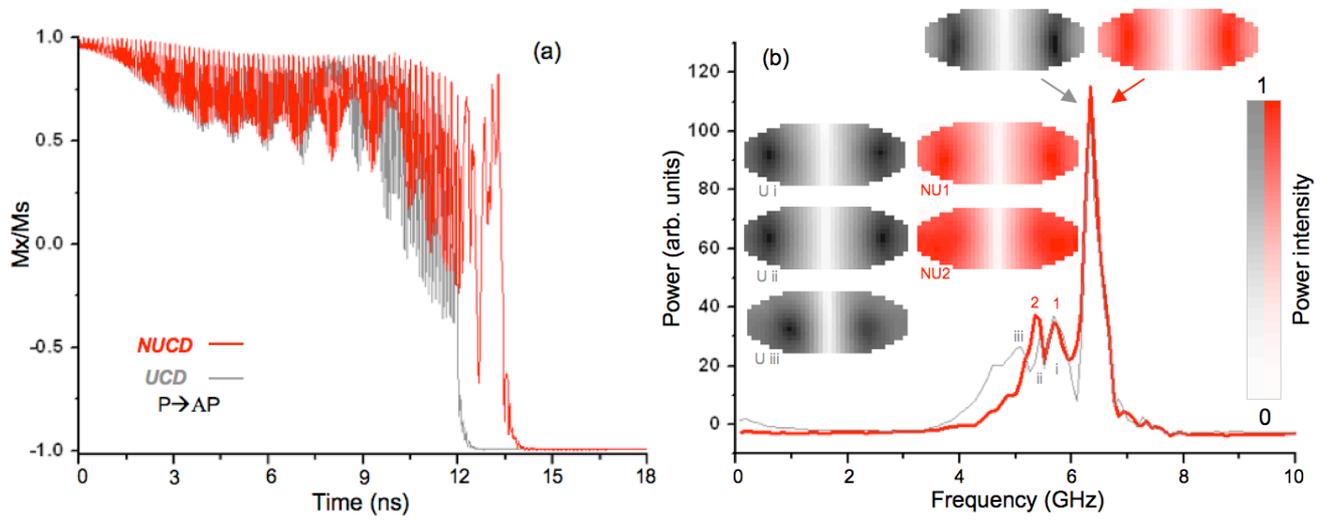

Fig. 4

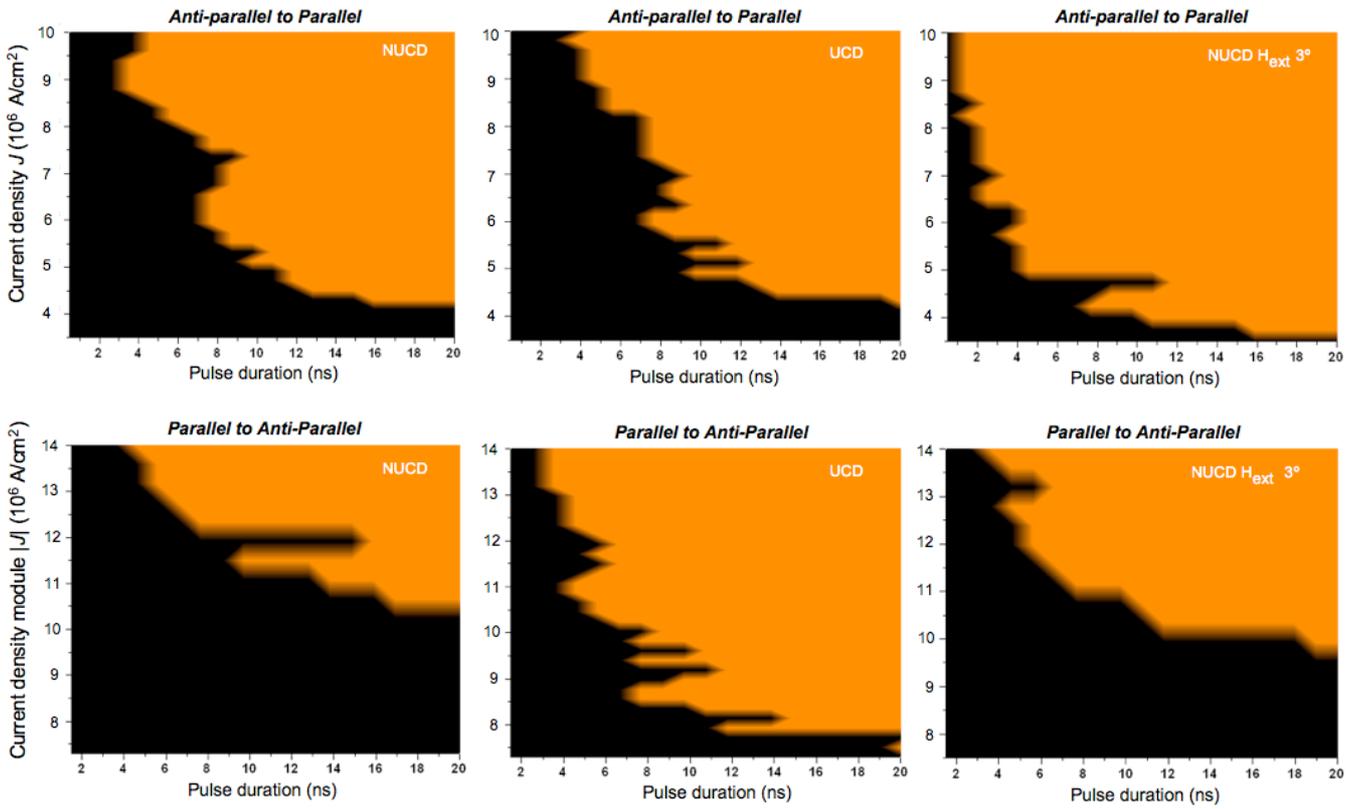

Fig. 5

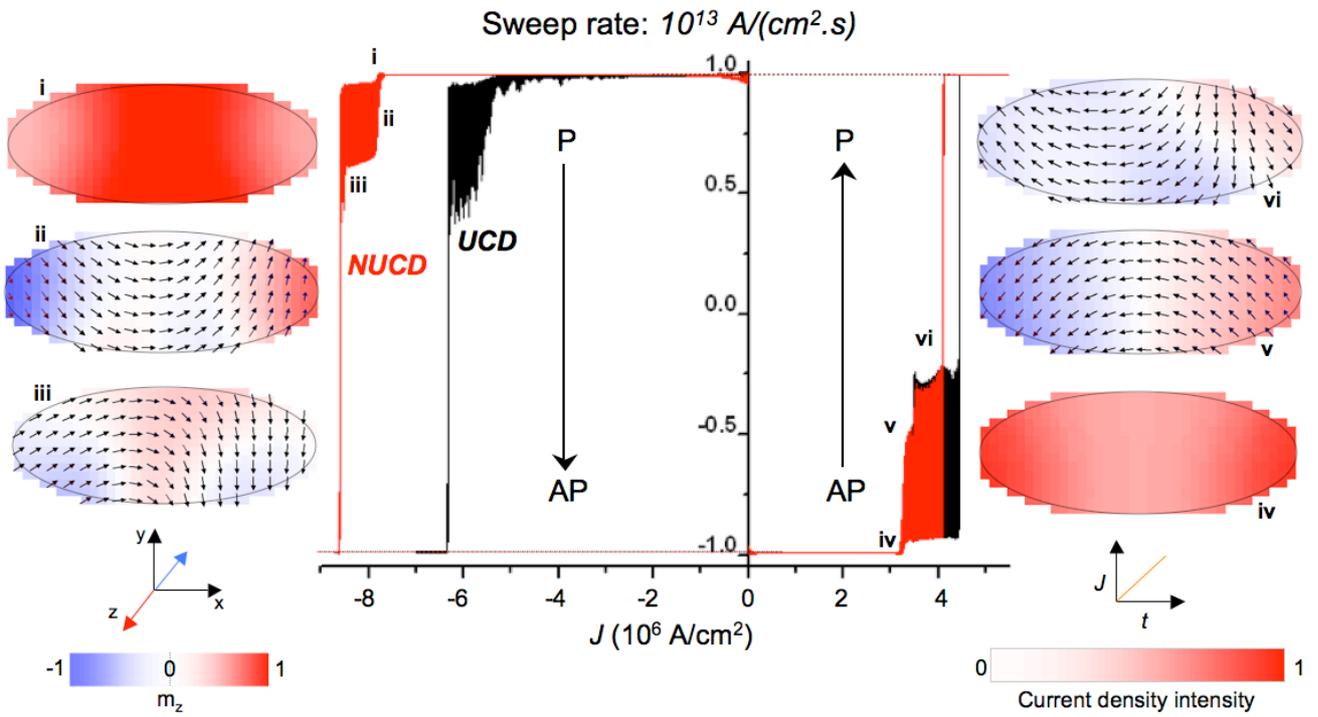

Fig. 6

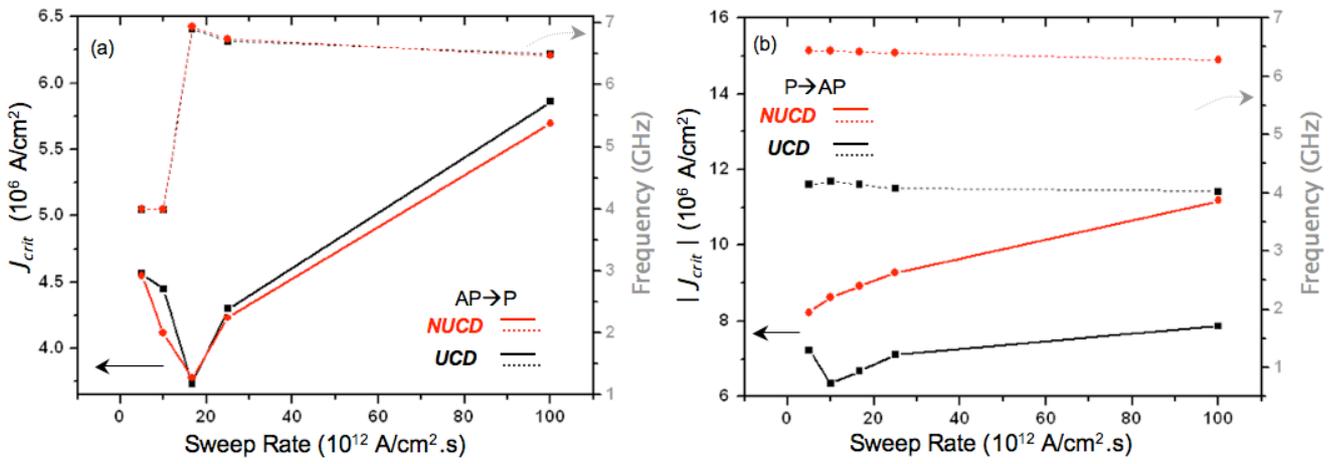

Fig. 7

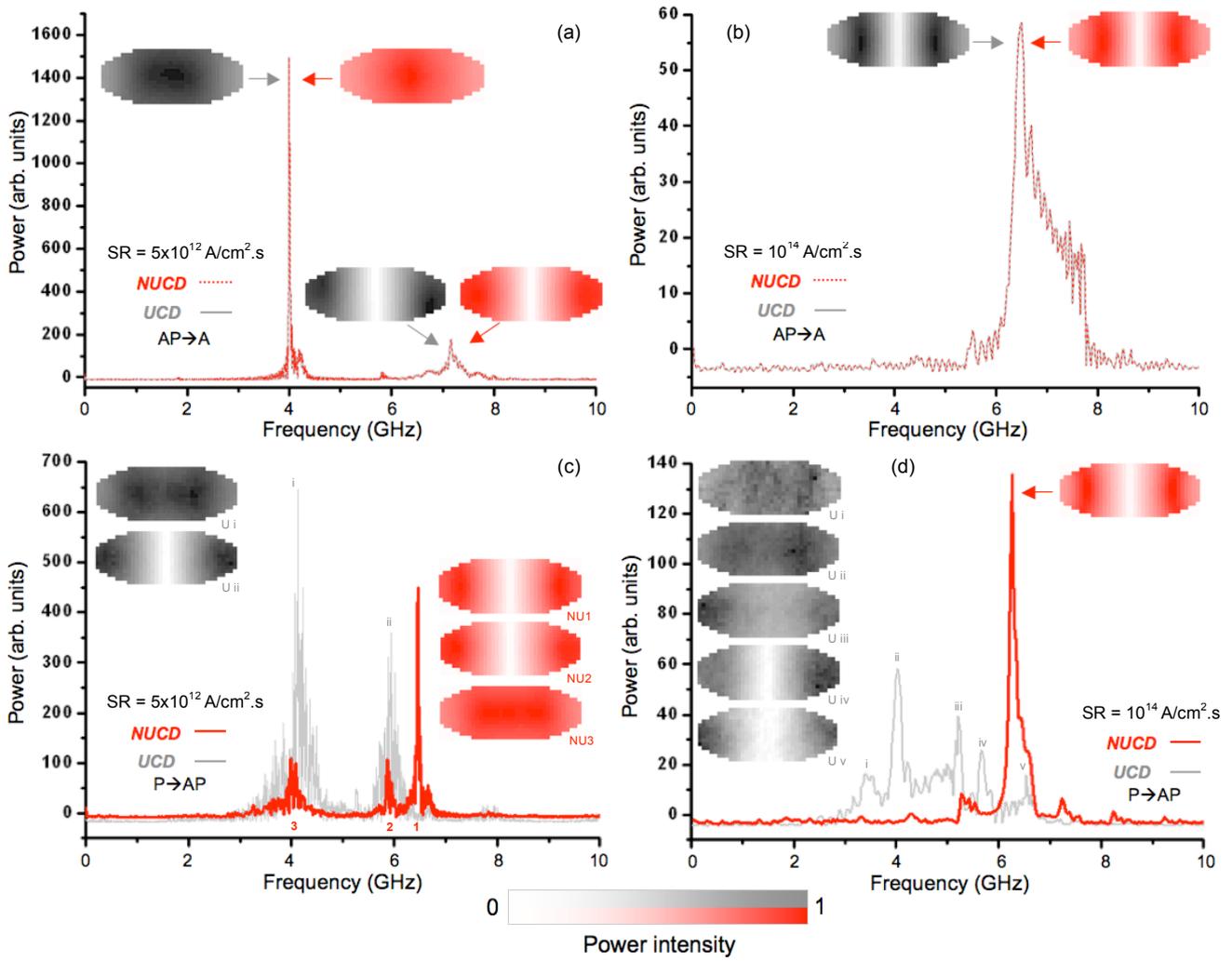

Fig. 8